\documentclass[twocolumn,pre,aps,floatfix]{revtex4}
\usepackage{graphics,graphicx,epsfig}
\usepackage{epsf,epstopdf,wrapfig}
\usepackage{amsmath,bm}

\usepackage{pgf,tikz}
\usetikzlibrary{arrows,automata}

\begin{document}

\title{Noise and information transmission in promoters with multiple internal states}
\author{Georg Rieckh and Ga\v{s}per Tka\v{c}ik}
\affiliation{Institute of Science and Technology Austria, Am Campus 1, A--3400 Klosterneuburg, Austria}

\date{\today}

\newcommand\startsupplement{%
    \makeatletter 
       \setcounter{figure}{0}
       \renewcommand{\thefigure}{S\arabic{figure}}
    \makeatother}

\begin{abstract}
Based on the  measurements of noise in gene expression performed during the last decade, it has become customary to think of gene regulation in terms of a two-state model, where the  promoter of a gene can stochastically switch between an ON and an OFF state. As experiments are becoming increasingly precise and the deviations from the two-state model start to be observable, we ask about the  experimental signatures of complex multi-state promoters, as well as the functional consequences of this additional complexity.
In detail, we (i) extend the calculations for noise in gene expression to promoters described by state transition diagrams with multiple states, (ii) systematically compute the experimentally accessible noise characteristics for these complex promoters, and (iii) use information theory to evaluate the channel capacities of  complex promoter architectures and compare  them to the baseline provided by the two-state model. We find that  adding internal states to the promoter generically decreases channel capacity, except in certain cases, three of which (cooperativity, dual-role regulation, promoter cycling) we analyze in detail. 
\end{abstract}

\maketitle

\section{Introduction}
Gene regulation -- the ability of cells to modulate the expression level of genes to match their current needs -- is crucial for survival. One important determinant of this process is the wiring diagram of the  regulatory network, specifying how environmental or internal signals are detected, propagated, and combined to orchestrate protein level changes \cite{LevineDavdidsson05}. Beyond the wiring diagram, the capacity of the network to reliably transmit information about signal variations is determined also by the strength of the network interactions (the ``numbers on the arrows'' \cite{RonenAlon2002}), the dynamics of the response, and the noise inherent to chemical processes happening at low copy numbers \cite{Elowitz2002, OzbudakvanOudenaarden2002, Paulsson04summing, RajvanOudenaarden2008rev}.

How do these factors combine to set the regulatory power of the cell? Information theory can provide a general measure of the limits to which a cell can reliably control its gene expression levels.
Especially in the context of developmental processes, where the precise establishment and readout of positional information has long been appreciated as crucial~\cite{Houchmandzadeh2002}, information theory can provide a quantitative proxy for the biological function of gene regulation \cite{TkacikWalczak2011}. This has led to theoretical predictions of optimal networks that maximize transmitted information given biophysical constraints \cite{ZivNemenmanWiggins2007, TkacikCallankBialekPRE2008, TkacikWalczakBialek2009, WalczakTkacikBialek2010, TkacikWalczak2011, TkacikWalczakBialek2012}, and hypotheses that certain biological networks might have evolved to maximize transmitted information \cite{TkacikBialek2008}. Some evidence for these ideas has been provided by recent high-precision measurements in the gap gene network of the fruit fly \cite{DubuisGregor2013}.
In parallel to this line of research, information theory has been used as a general and quantitative way to compare signal processing motifs \cite{TostevintenWolde2009, TostevintenWolde2010, CheoungLevchenko2011, deRondetenWolde2010, deRondetenWolde2012, deRondetenWolde2011, TostevintenWolde2012, MuglertenWolde2013, BowsherSwain2012, JostLevine2013, Hormoz2013, LevineMirny2007, ManciniWalczak2013}. Further theoretical work has demonstrated a relationship between the information capacity of an organism's regulatory circuits and its evolutionary fitness \cite{BialekTaylor, LeiblerRivoire, DonaldsonMatasciBergstromLachmann}.

Previously, information theoretic investigations  primarily examined the role of the regulatory network. Here we focus on the molecular level, i.e., on the events taking place at the regulatory regions of the DNA. Little is known about how the architecture of such microscopic events shapes information transfer in gene regulation. Yet it is precisely at these regulatory regions that the mapping from the ``inputs'' in the network wiring diagram into the corresponding ``output'' expression level is implemented by individual molecular interactions. In this bottleneck various physical sources of stochasticity -- such as the binding and diffusion of molecules \cite{BialekSetayeshgar05, BialekSetayeshgar08, GregorBialek2007}, and the discrete nature of chemical reactions \cite{vanKampen} -- must play an important role. In the simplest picture, gene expression is modulated through transcriptional regulation. This involves molecular events like the binding of transcription factors (TFs) to specific sites on the DNA, chemical events that facilitate or block TF binding (e.g., through chromatin modification), or events that are subsequently required to initiate transcription (e.g., the assembly and activation of the transcription machinery). 

While the exact sequence of molecular events at the regulatory regions often remains elusive (especially in eukaryotes), quantitative measurements have highlighted factors that contribute to the fidelity by which TFs can affect the expression of their target genes. These findings have been succinctly summarized by the so-called ``telegraph model'' of transcriptional regulation \cite{PeccoudYcart1995}: the two-state promoter switches stochastically between the states ``ON'' and ``OFF'', with switching rates dependent on the concentration(s) of the regulatory factor(s). This dependence can either be  biophysically motivated (e.g. by a thermodynamic model of TF binding to DNA), or it can be considered as purely phenomenological. The switching itself is independent of mRNA production, but determines the overall production rate. The production of mRNA molecules from one state is usually modeled as a Poisson process, with a first-order decay of messages; this is usually followed by a birth-death process in which proteins are translated from the messages. This two-state model is well-studied theoretically \cite{PeccoudYcart1995, IyerBiswasHayotJayaprakash2009, ShahrezaeiSwain2008, DobrzynskiBruggeman2009, KeplerElston2001, HuLevine2011, HuLevine2012} and has been used extensively to account for measurements of noise in gene expression \cite{SoGolding2011, TkacikGregorBialek2008, RaserOShea2004, Raj2006}. An increasing amount of information about molecular details has motivated extensions to this model by introducing  more than two states in specific systems  \cite{SanchezKondev2008,Gutierrez2009, Karmakar2010, CoulonGrandillonBeslon2010, SanchezKondev2011, ZhouZhang2012, ZhangChenZhou2012,Gutierrez2012}, and recent measurements of noise in gene expression provided some support for such complex regulatory schemes \cite{BlakeBalazsiCantorCollins2006, BlakeKaernCantorCollins2003, KandhaveluRibeiro2011, SuterSchiblerNaef2011}.

Here we address the general question of the functional effect of complex promoters with multiple internal states. How does the presence of multiple promoter states affect information transmission? Which promoter architectures transmit information more reliably when placed into a regulatory network? Under what conditions, if any, can multi-state promoters perform better than the two-state model?
To address these questions, we consider a wide spectrum of generic promoter models that can be treated mathematically as state transition diagrams; many molecular ``implementations'' could thus share the same underlying model.  When placed into a network, one must further specify which of the transitions are affected by concentrations of regulatory proteins, and which of the promoter states have nonzero expression rates. With this framework in hand, we derive the total noise in mRNA expression as a function of the induction level for all two- and three-state promoter models, and discuss how measurements of this function can be diagnostic of the underlying mechanism of regulation. To answer the main question of this paper -- namely if additional complexity at the promoter can lead to an improvement in controlling the output level of a gene -- we compute the information transmission from transcription factor concentrations to regulated protein expression levels through two- and three-state promoters. Finally, we analyze in detail three complex promoter architectures that outperform the two-state regulation.

\begin{figure*}
   \begin{center}
      \includegraphics*[width=6.5in]{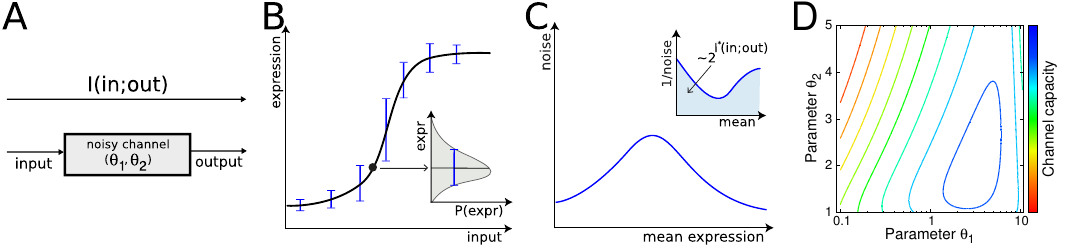}
      \caption{{\bf A genetic regulatory element as an information channel.}
  {\bf (A)}~Mutual information $I$  is a quantitative  measure of the signaling fidelity with which a genetic regulatory element maps inputs (e.g. TF concentrations) into the regulated expression levels. In this schematic example, the properties of the element are fully specified by two parameters $(\theta_1,\theta_2$; e.g., the switching rates between promoter states). {\bf (B)} In steady state, the input/output mapping can be summarized by the regulatory function $\bar{g}(k)$ (solid black line) for target protein expression (equivalently, $\bar{m}(k)$ for target mRNA expression, not shown); noise, $\sigma_g(k)$ (resp. $\sigma_m(k)$ for mRNA), induces fluctuations around this curve (inset and error bars on the regulatory function). {\bf (C)} The ``noise characteristic''  (noise vs. mean expression) is usually experimentally accessible for mRNA using \emph{in situ} hybridization methods and can reveal details about the promoter architecture. The maximal transmitted information (channel capacity $I^*$, see Eqs~(\ref{info2},\ref{info3})) is calculated from the area under the inverse noise curve for the target protein, $\sigma_g^{-1}(\bar{g})$ (inset). {\bf (D)} Channel capacity $I^*$ is, in this example case, maximized for a specific choice of parameters $\theta_1, \theta_2$ (blue peak).}
      \label{fig:setup}
   \end{center}
\end{figure*}

\section{Methods}

\subsection{Channel capacity as a measure of regulatory power}
We start by considering a genetic regulatory element -- e.g., a promoter or an enhancer -- as a communication channel, shown in Fig~\ref{fig:setup}A. As the concentrations of the relevant inputs (for example, transcription factors) change, the regulatory element responds by varying the rate of target gene expression. In steady state, the relationship between input $k$ and expression level of the regulated protein $g$ is often thought of as a ``regulatory function'' \cite{SettyAlon2003}. While attractive, the notion of a regulatory \emph{function} in a mathematical sense is perhaps misleading: gene regulation is a noisy process, and so for a fixed value of the input we have not one, but a distribution of different possible output expression levels, $P(g|k)$ (see Fig~\ref{fig:setup}B). When the noise is small, it is useful to think of a regulatory function as describing the \emph{average} expression level, $\bar{g}(k)=\int dg\, gP(g|k)$, and of the noise as inducing some random fluctuation around that average. The variance of these fluctuations, $\sigma^2_g(k)=\int dg\, (g-\bar{g}(k))^2P(g|k)$, is thus a measure of noise in the regulatory element; note that its magnitude depends on the input, $k$. 

The presence of noise puts a bound on how precisely changes in the input can be mapped into resulting expression levels on the output side -- or inversely, how much the cell can know about the input by observing the (noisy) outputs alone. In his seminal work on information theory \cite{Shannon48}, Shannon introduced a way to quantify this intuition by means of \emph{mutual information}, which is an assumption-free, positive scalar measure in bits, defined as
\begin{equation}
I(k;g)=\iint dk\, dg\ P(k) P(g|k) \log_2\left[\frac{P(g|k)}{P(g)}\right] \quad . \label{info1}
\end{equation}
In Eq~(\ref{info1}), $P(g|k)$ is a property of the regulatory element, which we will be computing below, while $P(k)$ is the distribution of inputs (e.g. TF concentrations) that regulate the expression; finally, $P(g)=\int dk\, P(g|k)P(k)$ is the resulting distribution of gene expression levels.
With $P(g|k)$ set by the properties of the regulatory element and the biophysics of the gene expression machinery, there exists an optimal choice for the distribution of inputs, $P^*(k)$, that maximizes the transmitted information. This maximal value, $I^*(k;g)=\mathrm{max}_{P(k)}I(k;g)$, also known as the \emph{channel capacity} \cite{CoverThomas}, summarizes in a single number the ``regulatory power'' intrinsic to the regulatory element \cite{TkacikBialek2008, TkacikCallankBialekPRE2008, TkacikWalczakBialek2009, WalczakTkacikBialek2010, TkacikWalczakBialek2012}.

Our goal is to compute the channel capacity between the (single) regulatory input and the target gene expression level for information flowing through various complex promoters. 
Under the assumption that noise is small and approximately gaussian for all levels of input, the complicated expression for the information transmission in Eq~(\ref{info1}) simplifies, and the channel capacity $I^*(k;g)$ can be computed analytically from the regulatory function, $\bar{g}(k)$, and the noise, $\sigma^2_g(k)$. The result is that \cite{TkacikBialek2008, TkacikCallankBialekPRE2008, TkacikWalczakBialek2009, WalczakTkacikBialek2010, TkacikWalczakBialek2012}
\begin{eqnarray}
I^*(k;g)&=&\log_2\frac{Z}{\sqrt{2\pi e}}\quad , \quad  \rm{with} \label{info2} \\
Z &=& \int_0^{k_{\rm max}}dk\frac{ |d\bar{g}/dk|}{\sigma_g(k)} = \int_{\bar{g}_{\rm min}}^{\bar{g}_{\rm max}}\frac{d\bar{g}}{\sigma_g(\bar{g})}\quad , \label{info3}
\end{eqnarray}
where in the last equality we changed the integration variables to express the result in terms of the average induction level, $\bar{g}$, using the regulatory function $\bar{g} = \bar{g}(k)$. This integral is graphically depicted in Fig~\ref{fig:setup}C (inset). Finally, we will use this to explore the dependence of $I^*(k;g)$ on parameters that define the promoter architecture (see Fig~\ref{fig:setup}D), looking for those arrangements that lead to large channel capacities and thus high regulatory power.

Information as a measure of regulatory power has a number of attractive mathematical properties (for review, see \cite{TkacikWalczak2011}); interpretation-wise, the crucial property is that it roughly counts (the logarithm of) the number of distinguishable levels of expression that are accessible by varying the input -- also taking into account the level of noise in the system. A capacity of~1~bit therefore suggests that the gene regulatory element could act as a binary switch with two distinguishable expression levels
; capacities smaller than~1~bit correspond to (biased) stochastic switching, while capacities higher than~1~bit support graded regulation. An increase of information by~1~bit means that the number of tunable and distinguishable levels of gene expression has roughly doubled~(!), implying that  changes of less than a bit are meaningful. Careful analyses of gene expression data for single-input single-output transcriptional regulation suggest that real capacities can exceed~1~bit \cite{TkacikBialek2008}. Increasing this number substantially beyond a few bits, however, necessitates very low levels of noise in gene regulation, requiring prohibitive numbers of signaling molecules \cite{TkacikCallankBialekPRE2008}.

\subsection{Multi-state promoters as state transition graphs}
\label{sec:framework}
\begin{figure}
  \begin{center}
      \includegraphics[width=\linewidth]{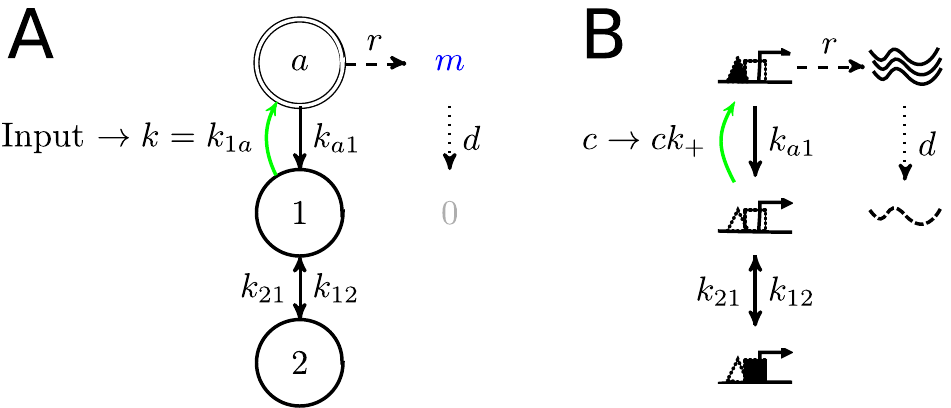}
      \caption{ {\bf Promoters as state transition graphs.}
  {\bf (A)}~A state transition graph for an example 3-state promoter. Active state $a$ (double circle) expresses mRNA $m$ at rate $r$, which are then degraded with rate $d$. Transition into $a$ (green arrow) is affected by the input that modulates rate $k=k_{1a}$. Stochastic transitions between promoter states $\{a,1,2\}$ are an important contribution to the noise, $\sigma_m(k)$.
  {\bf (B)} A possible mechanistic interpretation of the diagram in A: state 1 is an unoccupied promoter, state 2 is an inaccessible promoter (occupied by a nucleosome or repressor, black square). Transition to the active state (green arrow) is modulated by changing the concentration $c$ of activators (filled triangles) which bind their cognate site (empty triangles) at the promoter with the rate $ck_+$.}
      \label{fig:stategraph}
  \end{center}
\end{figure}

To study information transmission, we must first introduce the noise model in gene regulation, which consists of two components: \emph{(i)}, the generalization of the random telegraph model to multiple states, and \emph{(ii)}, the model for input noise that captures fluctuations in the number of regulatory molecules.
Starting with \emph{(i)}, we compute here the mean and variance for regulated mRNA levels, since these quantities are experimentally accessible when probing noise in gene expression. We assume that the system has reached steady state and that gene product degradation is the slowest timescale in the problem, i.e., that target mRNA or protein levels average over multiple state transitions of the promoter and that the resulting distributions of mRNA and protein are thus unimodal. While for protein levels these assumptions hold over a wide range of parameters and include many biologically relevant cases, there exist examples where promoter switching is very slow and the system would need to be treated with greater care (e.g., \cite{WalczakWolynes2005, Karmakar2010, IyerBiswasHayotJayaprakash2009}).

Let us represent the possible states of the promoter (and the transitions between them) by a state transition graph as in Fig~\ref{fig:stategraph}A. Gene regulation occurs when an input signal modifies one (or more) of the rates at which the promoter switches between its states.
To systematically analyze many promoter architectures, we choose not to endow from the start each graph with a mechanistic interpretation, which would map the abstract promoter states to various configurations of certain molecules on the regulatory regions of the DNA (as in Fig~\ref{fig:stategraph}B). This is because there might be numerous molecular realizations of the same abstract scheme, which will yield identical noise characteristics and identical information transmission. In Fig~\ref{fig:schemes} and Fig~S1, we discuss known examples related to different promoter architectures.

Given a specific promoter architecture, we would like to compute the first two moments of the mRNA distribution under the above assumptions. Here, we only sketch the method for the promoter in Fig~\ref{fig:stategraph}A; for a general description and details see Supplementary Information. We will denote the rate of mRNA production from the active state(s) by $r$ and its degradation rate by $d$. Let further $p_i$ be the fractional occupancy of state $i\in \{a,1,2\}$ and $k_{ij}$ the rate of transitioning from state $i$~to~$j,\ i\neq j$. Here, $a$ is the active state, and $1,2$ are the non-expressing states.
Equations~(\ref{eqn:overall}) and~(\ref{eqn:transmatrix})  then describe the behavior of the state occupancy and mRNA level $m$:
\begin{equation}
   \partial _t \bm p = \bm K \bm p + \bm \xi,\quad \partial _t m = r p_a - d m + \xi_m  ,\quad \text{with}\label{eqn:overall}
\end{equation}
\begin{equation}
   \bm K=\begin{bmatrix}
        -(k_{a1}+k_{a2}) & k_{1a} & k_{2a}  \\
        k_{a1} & -(k_{1a}+k_{12}) & k_{21}  \\
        k_{a2} & k_{12} & -(k_{2a}+k_{21})  \\
     \end{bmatrix}
  \quad , \label{eqn:transmatrix}
\end{equation}
and $\bm p=\left(p_a , p_1 , p_2\right)^T$;
$\bm \xi=\left(\xi_a, \xi_1, \xi_2\right)^T$ and $\xi_m$ are Langevin white-noise random forces \cite{vanKampen, Gillespie2000} (see Supplementary Information). In this setup it is easy to compute the mean and the variance in expression levels given a set of chosen rate constants.
Using the assumption of slow gene product degradation, $d\ll k_{ij}$, we can write the noise in a generic way:
\begin{eqnarray}
\sigma_m^2=
\bar m \left[ 1+\frac{r}{d}\ p_{\rm act} \cdot \Delta \right]  , \label{eqn:var1}
\end{eqnarray}
where $p_{\rm act}$ is the occupancy of the active states ($p_a$ or $p_a+p_b$), and the dimensionless expression for $\Delta$ depends on the promoter architecture and can be read out from Fig~\ref{fig:schemes}A for different promoter models. The expression for noise in Eq~(\ref{eqn:var1}) has two contributions. The first, where the variance is equal to the mean ($\sigma_m^2=\bar{m}+\dots$) is the ``output noise''  due to the birth-death production of single mRNA molecules (also called ``shot noise'' or ``Poisson noise''). The second contribution to the variance in  Eq~(\ref{eqn:var1}) is due to stochastic switching of the promoter between internal states, referred to as  the ``switching noise.'' This term does depend on the promoter architecture and has a more complicated functional form than being simply proportional to the mean. A first glance at the expressions for noise seems to imply that going from two to three promoter states can only increase the noise (and by Eq~(\ref{info3}) decrease information), since new, positive contributions appear in the expressions for $\sigma_m^2$; we will see that, nevertheless, transmitted information can increase for certain architectures.

\subsection{Input noise}
In addition to the noise sources internal to the regulatory mechanism, we also consider the propagation of fluctuations in the input, which will contribute to the observed variance in the gene expression level.
Can we say anything general about the transmission of input fluctuations through the genetic regulatory element? Consider, for instance, the modulated rate $k$ that depends on the concentration $c$ of some transcription factor, as in $k=k_+ c$, where $k_+$ is the association rate to the TF's binding site. Since the TF itself is expressed in a stochastic process, we could expect that there will be (at least) Poisson-like fluctuations in $c$ itself, such that $\sigma_c^2\propto c$; this will lead to an effective variance in $k$ that will be propagated to the  output variance in proportion to the ``susceptibility'' of the regulatory element, $\left(\partial \bar{g}/\partial k\right)^2$. Extrinsic noise would affect the regulatory element in an analogous way, as suggested in Ref~\cite{SwainSiggia02}. Independently of the noise origin, we can write
\begin{equation}
\sigma_g^2=\dots+v \left(\frac{\partial \bar{g}}{\partial k}\right)^2 k, \label{eqn:diffnoise}
\end{equation}
where $(\ldots)$ indicate output and switching terms from Eq~(\ref{eqn:var2}) and $v$ is the proportionality constant ($\sigma_k^2=vk$) that is related to the magnitude of the input fluctuations  and, possibly, their subsequent time averaging \cite{Paulsson04summing}.

Even if there were absolutely no fluctuations in the total concentration $c$ of transcription factor molecules in the cell (or the nucleus), the sole fact that they need to find the regulatory site by diffusion puts a lower bound on the variance of the \emph{local concentration} at the regulatory site. This diffusion limit, first formulated for the case of bacterial chemotaxis by Berg and Purcell \cite{BergPurcell1977}, has been subsequently derived for the general case of biochemical signaling \cite{BialekSetayeshgar05, BialekSetayeshgar08}: the lower bound on the variance in local concentration obeys $\sigma_c^2\propto c d'/ D \ell $, where $D$ is the diffusion constant of the TF molecules, $\ell$ is the linear size of the binding site, and $1/d'$ is the noise averaging time (here the lifetime of the gene product). Analyses of high-precision measurements in gene expression noise during early fruit fly development have shown that this diffusion noise represents a substantial contribution to the total \cite{GregorBialek2007, TkacikGregorBialek2008}. Thus, for this biophysical limit set by diffusion, we find yet again that the variance in the input is proportional to the input itself. This, in sum, demonstrates that Eq~(\ref{eqn:diffnoise}) can be used as a very generic model for diverse kinds of input noise.
To see which values the constant $v$ can take, note that $\sigma_k^2=k_+^2\sigma_c^2 \propto k_+^2 c  d' / D \ell=k k_+ d'/ D \ell$. As an example, consider diffusion-limited association, where $k_+=4\pi D \ell$ \cite{BergvonHippel1985}.
Depending on the accuracy and the geometry of the sensing mechanism we now get  different values for $\tilde{v}=v/d'$, but in general $\tilde{v}$ is expected to be of order unity. For example, the perfect absorbing sphere has $\sigma_c^2 = c d'/(4\pi D\ell) $ and therefore $\tilde v = 1$; the perfect monitoring sphere in the Berg--Purcell limit has $\sigma_c^2 = 3c d'/(5\pi D\ell) $ and therefore $\tilde v = 2.4$ \cite{EndresWingreen2008, BergPurcell1977}.

\section{Results}

\subsection{Experimentally accessible noise characteristics}
\label{sec:signatures}
\begin{figure*}
\begin{center}
  \includegraphics*[width=6.5in]{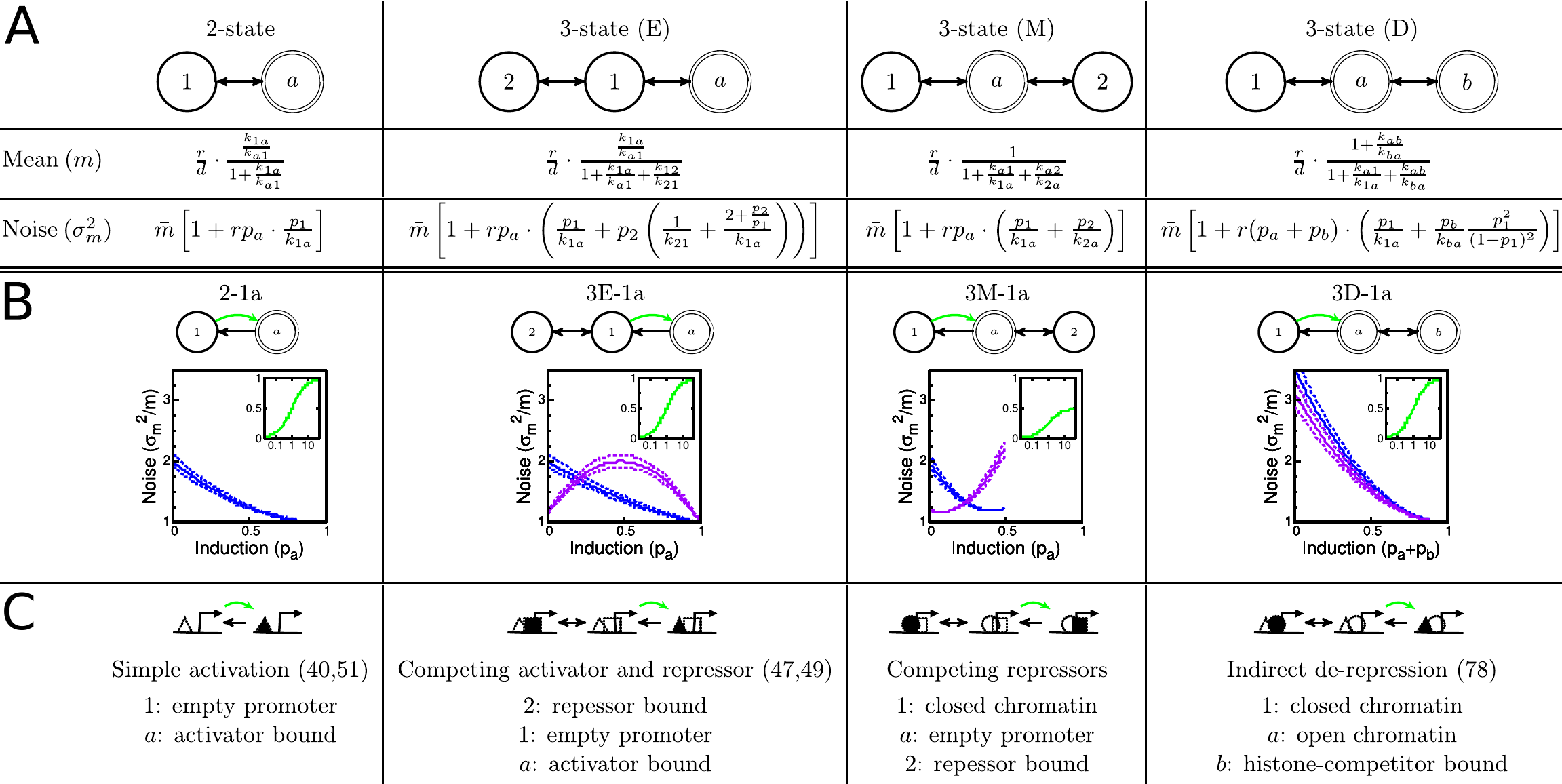}
  \caption{{\bf Mean expression level and noise for different promoter models.}
  {\bf (A)}~Expressions for the mean (first row, $\bar m$) and the variance (second row, $ \sigma_m^2$) of the mRNA distribution in steady state for different promoter architectures. In the limit of $k_{12}\to 0$ (resp. $k_{a2}, k_{ab}\to 0$), the expression for the two-state model is obtained from the models with three states. The names of the topologies indicate the position of the expressing state: E(nd), M(iddle), D(ouble).
  {\bf (B)}~Noise characteristics ($\bar{m}$ on $x$-axis vs the Fano factor, $\sigma_m^2/\bar m $, on $y$-axis)  for different promoter models. Here, in all models $k=k_{1a}$ is modulated (green arrow) to achieve different mean expression levels. For all rates (except $k$) equal to 1 (blue lines), the functional form of the noise characteristics is very similar. This remains true for a variation of $k_{a1}$ of $\pm10$\% (blue dashed lines). Making the rates in/out of the third state (state $2$ or $b$) slower by a factor of 5 (purple lines, dashed $\pm10$\%) yields qualitatively different results. Insets show the induction curve, $p_{\rm act}(k)$, where $k$ is the modulated rate. A full table and possible molecular interpretations of different promoter schemes are given in Fig~S1~\cite{VinuelasBeslonGandrillon2013, SanchezGelles2011, GarciaPhillips2012CellRep, YangKo2012, Earnest2013, Mirny2010, SegalWidom2009rev}.
{\bf (C)} Possible interpretations of the modulation schemes. Triangles represent activators, squares are repressors and circles are histones. The dotted shapes denote (empty) binding sites. Cited references use similar models.}
  \label{fig:schemes}
  \end{center}
\end{figure*}

Could complex promoter architectures be distinguished by their noise signatures, even in the easiest case where the input noise can be neglected (as is often assumed \cite{TkacikGregorBialek2008})?
The expressions for the noise presented in Fig~\ref{fig:schemes}A hold independently of which transition rate the input is modulating. We can specialize these results by choosing the \emph{modulation scheme}, that is, making one (or more) of the transition rates the regulated one. This allows us to construct the regulatory function (insets in Fig~\ref{fig:schemes}B). Additionally, we can also plot the noise (here shown as the Fano factor, $\sigma_m^2/\bar{m}$) as a function of the mean expression, $\bar{m}$, thus getting the \emph{noise characteristic} of every modulation scheme. These curves, shown in Fig~\ref{fig:schemes}B, are often accessible from experiments \cite{CareySegal2013, SoGolding2011}, even when the identity of the expressing state or the mechanism of modulation are unknown. We systematically organize our results in Fig~\ref{fig:schemes}B (for the case when $k=k_{1a}$ is modulated), and provide a full version in  Fig~S1; we also list four molecular schemes implementing these architectures in Fig~\ref{fig:schemes}C, while providing additional molecular implementations in Fig~S1. We emphasize that very different molecular mechanisms of regulation can be represented by the same architecture, resulting in the same mathematical analysis and information capacity.

Measured noise-vs-mean curves have been used to distinguish between various regulation models \cite{deRondeNemenman2009, CareySegal2013, SoGolding2011}. For this, two conditions have to be met \cite{TkacikGregorBialek2008, SanchezKondev2013rev}. First, it must be possible to access the full dynamic range of the gene expression in an experiment, and this sometimes seems hard to ensure. The second condition is that the input noise is not the dominant source of noise: input noise can mimic promoter switching noise and can, e.g.,  provide alternative explanations for noise measurements in \cite{SoGolding2011} that quantitatively fit the data (not shown).

Even if these conditions are met, it would be impossible to distinguish between certain promoter architectures (e.g., 2-a1 vs. 3E-a1) with this method, while some would require data of a very high quality to distinguish (e.g., activating 3E-1a vs. repressing 3E-12, see Fig~S1), at least in certain parameter regimes. On the other hand, there exist noise characteristics that can only be obtained with multiple states (e.g., 3M-1a).

One feature that can easily be extracted from the measured noise characteristics is the asymptotic induction: it can be equal to $1$ (e.g., in 2-1a), or bounded away from $1$ (e.g., in 3M-1a). While this distinction between architectures cannot be inferred from the shapes of the regulatory functions, the effect on the noise characteristics is unambiguous: in the case where the expressing state is never saturated, the Fano factor does not drop to the Poisson limit of 1 even at the highest expression levels (which seems to have been the case in Ref~\cite{SoGolding2011}).

Taken together, when the range of promoter architectures is extended beyond the two-state model, distinguishing between these architectures based on the noise characteristics seems possible only under restricted conditions, emphasizing the need for dynamical measurements that directly probe transition rates (e.g., \cite{SuterSchiblerNaef2011, GoldingCox2005}), or for the measurements of the full mRNA distribution (rather then only its second moment). We note that dynamic rates are often reported \emph{assuming} the two-state model, as they are inferred from the steady state noise measurements (e.g., \cite{SoGolding2011,Raj2006}), and only a few experiments probe the rates directly (e.g., \cite{GeertzMaerkl2012}); for a brief review of the rates and their typical magnitudes see Supplementary Information.

\subsection{Information transmission in simple gene regulatory elements}
\label{sec:formalism}

\begin{figure}[!h]
  \begin{center}
  \includegraphics[width=\linewidth]{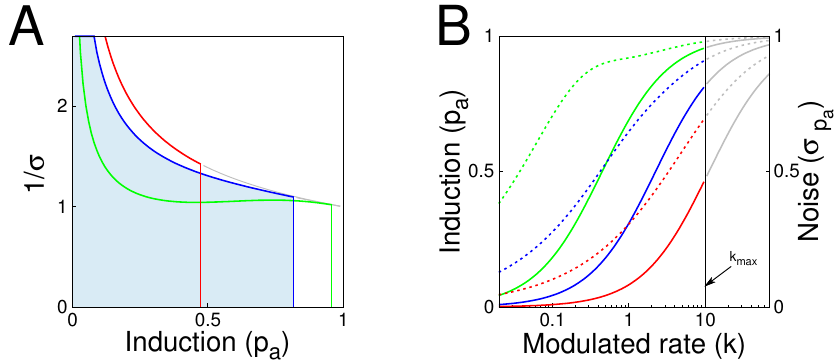}
  \caption{ {\bf Finding $k_-$ that optimizes information transmission in a two-state promoter.} The strength of the input noise is fixed at $v=2$ and the input dynamic range for $k$ is from 0 to $k_{\rm max}=10$. {\bf (A)} The integrand of Eq~(\ref{eqn:twostatez}) is shown for the optimal choice of $k_-$ (blue), and for two alternative $k_-$ values: a factor of 5 larger (red) or smaller (green) than the optimum. While increasing $k_-$ lowers the noise, it also decreases the integration limit, and vice versa for decreasing $k_-$. {\bf (B)} The effect on the regulatory function (solid, left axis) and the noise (dashed, right axis), of choosing different $k_-$ values. Optimal $k_-$ (blue curves) from (A) leads to a balance between the dynamic range in the mean response (the maximal achievable induction), and the magnitude of the noise. Higher $k_-$ (red curves), in contrast, lead to smaller noise, but fail to make use of the full dynamic range of the response. The gray part of the regulation curves cannot be accessed, since the input only ranges over $k\in [0,k_{\rm max}]$.}
  \label{fig:optimize}
  \end{center}
\end{figure}

{\bf Protein noise.}
In most cases the functional output of a genetic regulatory element is not the mRNA, but the translated protein.
Incorporating stochastic protein production into the noise model does not affect the functional form of the noise, but only rescales the magnitude of the noise terms. To see this, we let proteins be produced from mRNA at a rate $r'$ and degraded or diluted at $d'$, such that $d'\ll d$ is the slowest timescale in the problem. Then the mean protein expression level is $\bar{g}(k) = (r'/d') \bar{m}(k)$. The output and promoter switching noise contributions are affected differently, so that the protein level noise can be written as \cite{TkacikGregorBialek2008}:
\begin{equation}
\sigma_g^2=\bar{g}\left[(1+\nu) + \frac{r'}{d'} \ p_{\rm act} \cdot\Delta'\right] + (\mathrm{input\; noise})\ , \label{eqn:var2}
\end{equation}
where $\nu = r'/d$ is the \emph{burst size} (the average number of proteins translated from one mRNA molecule), $\Delta'=\frac{r}{d}\frac{d'}{d} \Delta$ and the other quantities are as defined in Eq~(\ref{eqn:var1}).

{\bf Information transmission in the two-state model.}
To establish the baseline against which to compare complex promoters, we look first at the two-state promoter (2-1a). Here the transition into the active state is modulated by TF concentration $c$ via $k=k_{1a}=k_+c$, as it would be in the simple case of a single TF  molecule binding to an activator site to turn on transcription. Adding together the noise contributions of Eqs~(\ref{eqn:var2},\ref{eqn:diffnoise}), we obtain our model for the total noise:
\begin{equation}
\sigma_g^2 = \bar g \left[ (1+\nu)+ \frac{rr'}{d k_{a1}}(1-p_a)^2 +  \frac{v}{k_{a1}} \frac{rr'}{dd'}(1-p_a)^3 \right]  . \label{nn1}
\end{equation}

To compute the corresponding channel capacity, we use Eq~(\ref{info3}) with the noise given by Eq~(\ref{nn1}):
\begin{eqnarray}
Z&=&\int_{\bar{g}_{\rm min}}^{\bar{g}_{\rm max}} \frac{d\bar{g} }{\sigma_g(\bar{g})}=\sqrt{N_{\rm max}}\int_0^{p_a^{\rm max}}dp_a \times  \nonumber \\
&\times& p_a^{-1/2} \left[1 + \frac{1}{\tilde k_{-}}(1-p_a)^2+\frac{\tilde{v}}{\tilde k_{-}}(1-p_a)^3\right]^{-1/2} \label{eqn:twostatez}\\
&=&\sqrt{N_{\rm max}} Z_0 .
\end{eqnarray}
Here, $N_{\rm max} = (rr')/[(dd')(1+\nu)]$ and $\tilde k_{-}=k_{a1}/(d'N_{\rm max})$ is the dimensionless combination of parameters related to the off-rate for the TF dissociation from the binding site.
$N_{\rm max}$ can be interpreted as the  number of \emph{independently} produced output molecules when the promoter is fully induced \cite{TkacikWalczakBialek2009, TkacikWalczakBialek2012, WalczakTkacikBialek2010}.
In the case where mRNA transcription is the limiting step for protein synthesis, $N_{\rm max}$ corresponds to the maximal average number of mRNA synthesized during a protein lifetime: $N_{\rm max}=r/d' \cdot \nu/(1+\nu) \approx r/d'$.
With this choice of parameters, $N_{\rm max}$ affects $Z$ multiplicatively and thus simply adds a constant offset to the channel capacity $I^*$ [see Eq~(\ref{info2})] without affecting the parameter values that maximize capacity. In what follows we therefore disregard this additive offset, and examine in detail only $I^*=\log_2 Z_0$. We also only use dimensionless quantities (as above, e.g., the rates are expressed in units of $d'$), but leave out the tilde symbols for clarity.

{\bf Optimizing information transmission.} What parameters maximize the capacity of the two-state promoter 2-1a given by Eq~(\ref{eqn:twostatez})? Given that the dynamic range of input (e.g., TF concentration) is limited  \cite{TkacikCallankBialekPRE2008, TkacikWalczakBialek2009, TkacikWalczakBialek2012, WalczakTkacikBialek2010},  $k\in[0,k_{\rm max}]$, and given a choice of $v$ that determines the type and magnitude of input noise, the channel capacity $I^*$ for the two-state promoter only depends non-trivially on the choice of a single parameter, $k_-$ 
Figure~\ref{fig:optimize} shows the tradeoff that leads to the emergence of a well-defined optimal value for $k_-^*$: at a fixed dynamic range for the input, $k\in[0,k_{\rm max}]$, the information-maximizing solution chooses $k_-^*$ that balances the strength of binding (such that the dynamic range of expression is large), while simultaneously keeping the noise as low as possible. If this abstract promoter model were interpreted in mechanistic terms where a  TF binds to activate gene expression, then choosing the optimal $k_-$ would amount to choosing the optimal value for the dissociation constant of our TF; importantly, the existence of such a nontrivial optimum indicates that, at least in an information-theoretic sense, the best binding is not the tightest one \cite{LiBergElf09, GronlundLotstedtElf2013, TkacikCallankBialekPRE2008, TkacikWalczakBialek2009, WalczakTkacikBialek2010, TkacikWalczakBialek2012, LevineMirny2007}. This tradeoff between noise and dynamic range of outputs (also called ``plasticity'') has also been noticed in other contexts \cite{Lehner2010, BajicPoyatos2012}.

\begin{figure}[!h]
  \begin{center}
\includegraphics[width=\linewidth]{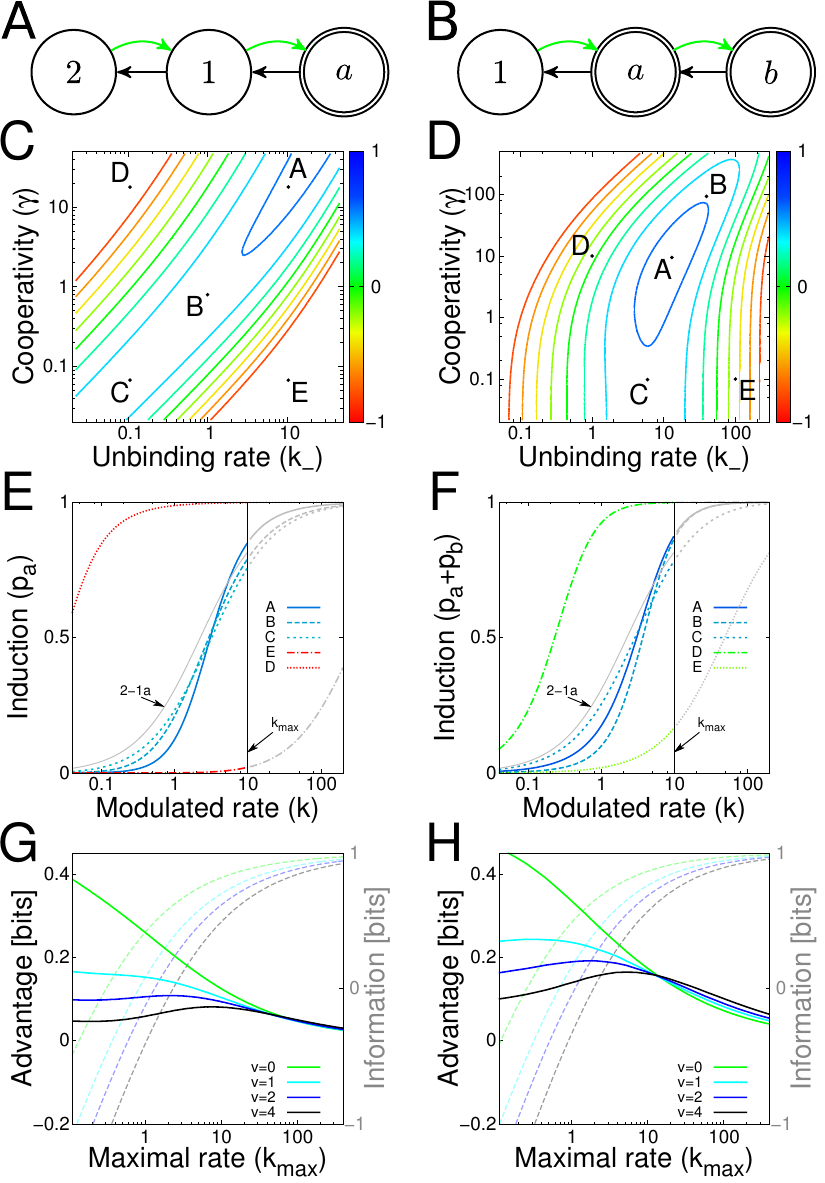}
  \caption{ {\bf Improving information transmission via cooperativity.}
  {\bf (A,B)} The state transition diagram for the AND-architecture three-state promoter (at left, one active state) and OR-architecture (at right, two active states).
  {\bf (C, D)} The information planes showing the channel capacity (color code) for various combinations of $k_-$ and $\gamma$ at a fixed maximum allowed $k_{\rm max}=10$, for the AND- and OR-architectures and $ v =2$.  
  {\bf (E, F)} The regulatory functions of various models selected from parameters denoted by dots in the information planes in (B, C). Solutions that maximize the information denoted with a solid blue line. Colors indicate the channel capacities. The gray part of the regulation curves cannot be accessed, since the input only ranges over $k\in [0,k_{\rm max}]$. For comparison, we also plot the regulation curve of the 2-1a scheme (with its optimal $k_-^*$).
 {\bf (G, H)} Channel capacity $I^*_3$ (dashed lines, axes at right) and the advantage of the multi-state scheme over the best two-state promoter, $I^*_3-I^*_2$ (solid lines, axes at left), of the AND and OR models, as a function of the maximal input range, $k_{\rm max}$, and the strength of the input noise ($ v$, color).  Since there is no globally optimal choice for $\gamma$ for the AND-architecture, we fix $\gamma =10$ in (G), and optimize only over $k_-$ values.
}
  \label{fig:coop}
  \end{center}
\end{figure}
\subsection{Improving information transmission with multi-state promoters}
\label{sec:improving}
We would like to know if complex promoter architectures can outperform the two-state model in terms of channel capacity. To this end, we have examined the full range of three-state promoters, summarized in Fig~S1, and found that generally -- as long as only one transition is modulated and only one state is active -- extra promoter states lead to a decrease in the channel capacity relative to two-state regulation. However, by relaxing these assumptions, architectures that outperform two-state promoters can be found.

{\bf Cooperative regulation.} The first such pair of architectures is illustrated in Fig~\ref{fig:coop}A and B: three-state promoters with one (or two) expressing states, where \emph{two} transitions into the expressing states are simultaneously modulated by the input. A possible molecular interpretation of these promoter state diagrams is an AND-architecture cooperative binding for the model with one expressing state, and an OR-architecture cooperative activation for the model with two expressing states. In case of an AND-architecture, a TF molecule hops onto the empty promoter (state 2) with rate $2k$ (since there are two empty binding site), while a second molecule can hop on with rate $\rho k$ (called ``recruitment'' if $\rho>1$), bringing the promoter into the active state. The first of two bound TF molecules falls off with rate $2 \gamma^{-1} k_-$ (called ``cooperativity'' if $\gamma > 1$), bringing the promoter back to state 1, and ultimately, the last TF molecule can fall off with rate $k_-$. The dynamics are now described (cf.~Eq~(\ref{eqn:transmatrix})) by the matrix
\begin{equation}
\bm K = 
  \begin{bmatrix}
  -(2\gamma^{-1} k_-) & \rho k & 0 \\
  2\gamma^{-1} k_- & -(k_-+\rho k) & 2 k \\
  0   & k_- & -(2 k) 
 \end{bmatrix}\ ,
\end{equation}
and $\bm p = \left( p_a, p_1, p_2\right)^T$, resp. $\bm p = \left(p_b, p_a, p_1\right)^T$.
To compute the noise, we can use the solutions for the generic three-state model 3E from Fig~\ref{fig:schemes}A by making the following substitutions: $k_{a1} = 2\gamma^{-1}k_-, \quad k_{1a} = \rho k, \quad k_{12}= k_-, \quad k_{21} = 2 k$.

To simplify our exploration of the parameter space, we choose $\rho=1$ (i.e., no recruitment), but keep $k_-$ (unbinding rate) and $\gamma$ (cooperativity) as free parameters; the modulated rate $k$ is proportional to the concentration of TF molecules and is allowed to range from $k\in [0, k_{\rm max}]$. For every choice of $(k_-,\gamma)$, we computed the regulatory function and the noise, and used these to compute the capacity, $I^*(k;g)$, using Eqs~(\ref{info2}, \ref{info3}). This information is shown in Fig~\ref{fig:coop}C and D for the AND- and OR-architecture, respectively. 

In the case of an AND-architecture, where both molecules of the TF have to bind for the promoter to express, there is a ridge of optimal solutions: as we move along the ridge in the direction of increasing information, cooperativity is increased and thus the doubly-occupied state is stabilized, while the unbinding rate increases as well. This means that the occupancy of state 1 becomes negligible, and the regulation function becomes ever steeper, as is clear from Fig~\ref{fig:coop}E, while maintaining the same effective dissociation constant (the input $k=k_{1/2}$ at which the promoter is half induced, i.e., $p_a(k_{1/2})=0.5$). In this limit, the shape of the regulation function must approach a Hill function with the Hill coefficient of 2, $p_a(k)=k^2/(k^2+k_{1/2}^2)$. Surprisingly, information maximization favors weak affinity of \emph{individual} TF molecules to the DNA, accompanied by strong cooperativity between these molecules.
The OR-architecture portrays a different picture: here, the maximum of information is well-defined for a particular combination of parameters $(k_-, \gamma)$, as shown in Fig~\ref{fig:coop}D. As $\gamma \to 0$ (increasing destabilization for $\gamma<1$), the second active state ($b$) is never occupied, and the model reverts to a two-state model. 

For both architectures we can assess the advantage of the three-state model relative to the optimal two-state promoter. Figures~\ref{fig:coop}E and F show the information of the optimal solutions as a function of the input noise magnitude as well as the input range, $k_{\rm max}$. As expected, the information increases as a function of $k_{\rm max}$ since the influence of input and switching noise can be made smaller with more input molecules. This increase saturates at high $k_{\rm max}$ because output noise becomes limiting to the information transmission -- this is why the capacity curves converge to the same maximum, the $v=0$ curve that lacks the input noise altogether. The advantage (increase in capacity) of the three-state models relative to the two-state promoter is positive for any combination of parameters $k_{\rm max}$ and $v$. It is interesting to note that increasing $k_{\rm max}$ and decreasing $v$ have very similar effects on channel capacity, since both drive the system to a regime where the limiting factor is the output noise.
\begin{figure}
  \begin{center}
\includegraphics*[width=3.25in]{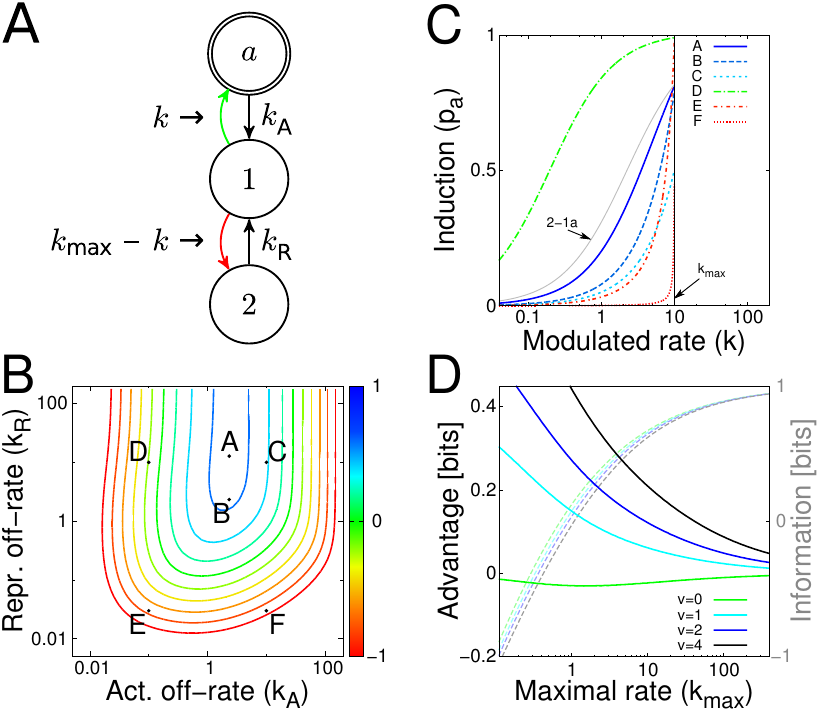}
  \caption{{\bf Improving information transmission via dual-role regulation.} {\bf (A)} The signal $k$ increases the concentration of TFs in the activator role that favor the transition (green) to the expressing state, while simultaneously decreasing the rate of switching (red) into the inactive state  with the repressor bound. {\bf (B)} Channel capacity (color) as a function of off-rates $(k_A, k_R)$ shows a peak at A. $k_{\rm max}=10$ and $v=4$. {\bf (C)} The regulatory functions for the optimal solution A (solid blue line) and other example points (B-E) from the information plane, show that this architecture can access a rich range of response steepnesses and induction thresholds. For comparison, we also plot the regulation curve of the scheme 2-1a (with its optimal $k_-^*$). {\bf (D)} The channel capacity (dashed line) and information advantage over the optimal two-state architecture (solid line), as a function of $k_{\rm max}$.}
  \label{fig:dual}
  \end{center}
\end{figure}

{\bf Regulation with dual-role TFs.} 
In the second architecture that we consider a transcription factor can switch its role from repressor to activator, depending on the covalent TF modification state or formation of a complex with specific co-factors. A well-studied example is in Hedgehog (Hh) signaling, where the TF Gli acts as a repressor when Hh is low, or as an activator when Hh is high \cite{Parker2011, White2012, MullerBasler2000}. Figure~\ref{fig:dual}A shows a possible dual-role signaling scheme where the total concentration of dual-role TFs is fixed (at $k_{\rm max}$), but the signal modulates the fraction of these TFs that play the activator role ($k$) and the remaining fraction that act as repressors ($k_{\rm max}-k$), which compete for the same binding site. The channel capacity of this motif is depicted in Fig~\ref{fig:dual}B as a function of promoter parameters $k_A$ and $k_R$, showing that a globally optimal setting (denoted ``A'') exists for these parameters; with these parameters, the input/output function, shown in Fig~\ref{fig:dual}C, is much steeper than what could be achieved with the best two-state promoter, and that is true despite the fact that the molecular implementation of this architecture uses only a single binding site. The ability to access such steep regulatory curves allows this architecture to position the mid-point of induction at higher inputs $k$, thus escaping the detrimental effects of the input noise at low $k$, while still being able to induce almost completely (i.e., make use of the full dynamic range of outputs) as the input varies from 0 to $k_{\rm max}$. This is how the dual-role regulation can escape the tradeoff faced by the two-state model 2-1a (shown in Fig~\ref{fig:optimize}). Sharper transition at higher input would lead us to expect that the advantage of this architecture over the two-state model is most pronounced when input noise is dominant (small $k_{\rm max}$, large $ v$), which is indeed the case, as shown in Fig~\ref{fig:dual}D.

\begin{figure}
  \begin{center}
   \includegraphics*[width=3.25in]{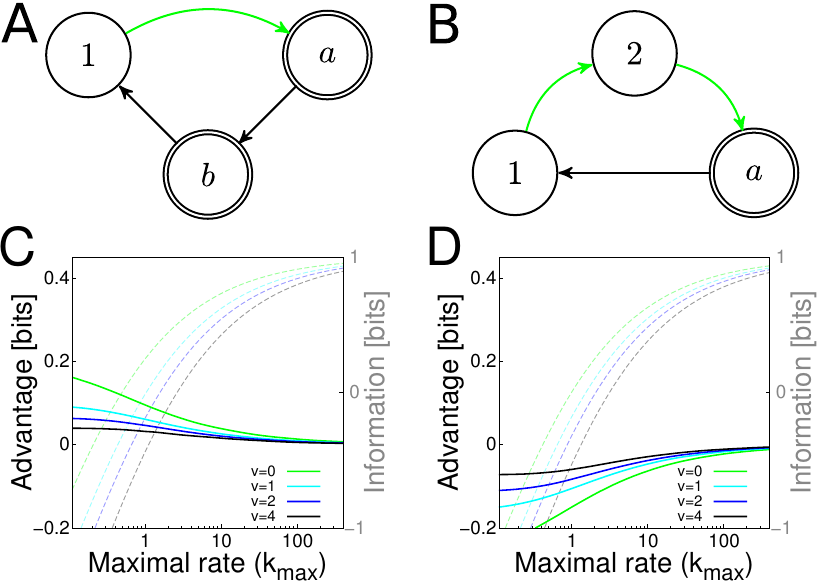}
  \caption{{\bf Improving information transmission via cycling.} {\bf (A)} Promoter cycles through two active states ($a,b$) expressing at identical production rates  before returning to inactive state ($1$), from which the transition rate back into the active state (green) is modulated. For each value of $k_{\rm max}$, we look for the optimal choice of $k_{ab}=k_{b1}$. The information and advantage relative to the optimal two-state model is shown in {\bf (C)}. {\bf (B,D)} A similar architecture where turning the gene on is a multi-step regulated process. This architecture always underperforms the optimal two-state model, indicated by a negative value of the advantage for all choices of $v$ and $k_{\rm max}$.}
  \label{fig:cycling}
  \end{center}
\end{figure}

{\bf Promotor cycling.}
In the last architecture considered here, promoters ``cycle'' through a sequence of states in a way that does not obey detailed balance, e.g., when state transitions involve expenditure of energy during irreversible reaction steps. In the scheme shown in Fig~\ref{fig:cycling}A, the regulated transition puts the promoter into an active state $a$; before decaying to an inactive state, the promoter must transition through another active state $b$. Effectively, this scheme is similar to the two-state model in which the decay from the active state is not first-order with exponentially distributed transition times, but rather with transition times that have a sharper peak. The benefits of this architecture are maximized when the transition rates from both active states are equal. While it always outperforms the optimal two-state model, the largest advantage is achievable for small $k_{\rm max}$. At large $k_{\rm max}$ the advantage tends to zero: this is because the optimal off-rates are high, causing the dwell times in the expressing states to be short. In this regime the gamma distribution of dwell times (in a three-state model) differs little from the exponential distribution (in a two-state model).  Note that this model would not yield any information advantage if the state transitions were reversible.

Figures~\ref{fig:cycling}C, D show that irreversible transitions alone do not generate an information advantage: a promoter that needs to transition between two inactive states ($1, 2$) to reach a single expressing state $a$ from which it exits in a first-order transition, is always at a loss compared to a two-state promoter. This is because here the effective transition rate to the active state in the equivalent two-state model is lower (since an intermediate state must be traversed to induce),  necessitating the use of a lower off-rate $k_{a1}$, which in turn leads to higher switching noise. 

It is interesting to note that recent experimental data on eukaryotic transcription seem to favor models in which the distribution of exit times from the expressing state is exponential, while the distribution of times from the inactive into the active state is not \cite{SuterSchiblerNaef2011}, pointing to the seemingly underperforming architecture of Fig~\ref{fig:cycling}B. The three-state promoter suggested here is probably an oversimplified model of reality, yet it nevertheless makes sense to ask why irreversible transitions through multiple states are needed to switch on the transcription of eukaryotic genes \cite{Larson2011rev,YosefRegevReview2011} and how this observation can be reconciled with the lower regulatory power of such architectures. This is the topic of our ongoing research.

\section{Discussion}
When studying noise in gene regulation one is usually restricted to the use of phenomenological models, rather than a fully detailed biochemical reaction scheme.
 Simpler models, such as those studied in this paper, also allow us to decouple questions of mechanistic interpretation from the questions of functional consequences. Here, we extended a well-known functional two-state model of gene expression to multiple internal states. We introduced  state transition graphs to model the ``decision logic'' by which  changes in the concentrations of regulatory proteins drive the switching of our genes between various states of expression. This abstract language allowed us to systematically organize and explore non-equivalent three-state promoters. The advantage of this approach is that many microscopically distinct regulatory schemes can be collapsed into equivalent classes sharing identical state transition graphs and identical information transmission properties.

The functional description of multi-state promoters confers two separate benefits. First, it is able to generate measurable predictions, such as the noise vs mean induction curve. Existing experimental and theoretical work using the two-state model has  demonstrated how the measurements of noise constrain the space of promoter models \cite{SoGolding2011}, how the theory establishes the ``vocabulary'' by which various measured promoters can be classified and compared to each other \cite{ZenklusenSinger2008}, and how useful a baseline mathematical model can be in establishing quantitative signatures of deviation which, when observed, must lead to minimal model revisions able to accommodate new data \cite{SuterSchiblerNaef2011}. 
Alternative complex promoters presented here could explain existing data better either because of the inclusion of additional states (c.f.~\cite{SanchezKondev2011}), or because we also included and analyzed the effects of input (diffusive) noise, which can mimic the effects of promoter switching noise but is often neglected \cite{TkacikGregorBialek2008}. As a caveat, it appears that in many cases discriminating between promoter architectures based on the noise characteristics alone would be very difficult, and thus dynamical measurements would be necessary.

The second benefit of our approach is to provide a convenient framework for assessing the functional impact of noise in gene regulation, as measured by the mutual information between the inputs and the gene expression level. We were interested in the question whether multi-state promoters can, at least in principle, perform better than the simple ON/OFF two-state model. We find that generically, i.e., for all three-state models where one state is expressing and only one transition is modulated by the input, the multi-state promoters underperform the two state model. Higher information transmission can be achieved when these conditions are violated, and biological examples for such violations can be found.
For example, we find that a multi-state promoter with cooperativity has a higher channel capacity than the best comparable two-state promoter, even when promoter switching noise is taken into account (c.f.  \cite{TkacikWalczakBialek2009}). Dual-regulation yields surprisingly high benefits, which are largest when input noise is high. In the context of metazoan development where the concentrations of the morphogen molecules can be in the nanomolar range and the input noise is therefore high \cite{GregorBialek2007}, the need to establish sharp spatial domains of downstream gene expression (as observed, \cite{DessaudBriscoe2008}) might have favored such dual-role promoter architectures. Lastly, we considered the simplest ideas for a promoter with irreversible transitions and have shown that they can lead to an increase in information transmission by sharpening the distribution of exit times from the expressing state \cite{PedrazaPaulsson2008}. 

The main conclusion of this article -- namely that channel capacity can be increased by particular complex promoters -- is testable in dedicated experiments. One could start with a simple regulatory scheme in a synthetic system and then by careful manipulation gradually introduce the possibility of additional states (e.g., by introducing more binding sites), using promoter sequences which show weaker binding for individual molecules yet allow for stronger cooperative interaction. In both the simple and complex system one could then measure the noise behavior for various input levels. Information theoretic analysis of the resulting data could be used to judge if the design of higher complexity, while perhaps noisier by some other measure, is capable of transmitting more information, as predicted.

The list of multi-state promoters that can outperform the two-state regulation and for which examples in nature could be found is potentially much longer and could include combinations of features described in this article. Rather than trying to find more examples, we should perhaps ask about the fundamentally different mechanisms and constraints that our analysis did not consider. In all cases that we analyzed the largest difference between the two- and three-state models was at low $k_{\rm max}$. This makes sense: at high $k_{\rm max}$ the dominant source of noise is the (bursty) Poisson production of gene products, which is the same regardless of the promoter architecture, while at low inputs, the input fluctuations filter through the promoter in ways that depend on its architecture. What other tricks could biology use to cope with input noise? By expending energy to keep the system out of equilibrium, one could design robust reaction schemes where, for example, the binding of a regulatory protein leads (almost) deterministically to some tightly controlled response cycle, perhaps evading the diffusion noise limit \cite{AquinoEndres2010} and increasing information transmission. At the same time, cells might be confronted by sources of stochasticity we did not discuss here, for example, due to cross-talk from spurious binding of non-cognate regulators. Finally, cells need to not only \emph{transmit} information through their regulatory elements, but actually perform \emph{computations}, that is, combine various inputs into a single output, thereby potentially discarding information. A challenging question for the future is thus about extending the information-theoretic framework to these other cases of interest.

\section{Acknowledgments}
We thank W Bialek, T Gregor, I Golding, M Draveck\'a, K Mitosch, G Chevereau, J Briscoe, and J Kondev for helpful discussions.


\appendix
\startsupplement

\begin{figure*}
  \begin{center}
  \includegraphics*[width=6.5in]{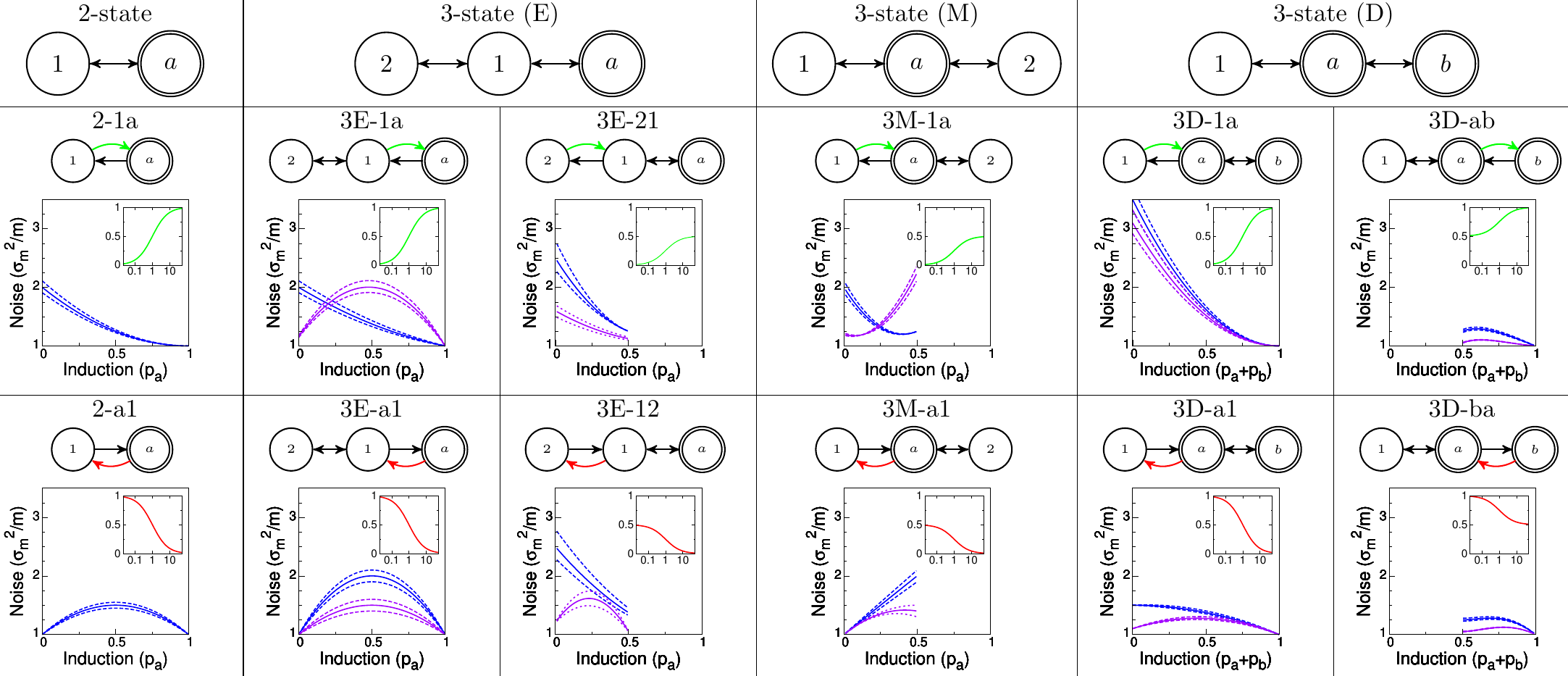}
  \caption{{\bf Promoter architectures and interpretations.}
  {\bf Scheme 2-1a:} (i) State $a$ is the empty promoter (available for transcription), and state $1$ corresponds to it being occupied by a repressing protein (e.g. a specific TF or a nucleosome). Mechanisms that change the rate of switching from state $1$ into $a$ (i.e. scheme 2-1a) are well documented for eukaryotic cells \cite{VinuelasBeslonGandrillon2013, CareySegal2013}. (ii) Simple activation where state $1$ is the empty promoter and state $a$ has an activating TF bound to it; changing TF concentration modulates $k_{1a}$ \cite{KeplerElston2001, SanchezKondev2011}.
  {\bf Scheme 2-a1:} The bacterial \emph{lac}-promoter, where the binding of a specific TF represses expression. A change in the concentration of TFs now corresponds to modulating the rate $k_{a1}$ and the rate $k_{1a}$ is determined by the interaction energy between TF and its binding site \cite{SanchezGelles2011, GarciaPhillips2012CellRep}.
  {\bf Scheme 3E:} (i) A promoter that has overlapping binding sites for both, an activator and a repressor (dual regulation) \cite{Karmakar2010, SanchezKondev2008, YangKo2012}. Changing the concentration of the activator (repressor) leads to a change in rate $k_{1a}$ ($k_{12}$), which means it can be modeled as a 3E-1a (3E-12) scheme. (ii) Eukaryotic promoters with a TATA~box can be modeled as 3E-21 \cite{SanchezKondev2013rev}. Here, state $2$ is the DNA in a state not available for transcription (closed chromatin), state $1$ is the conformation where the promoter (and the TATA~box) is exposed and state $a$ corresponds to the active configuration of the DNA with the pre-initiation complex assembled. Changing the concentration of chromatin remodelers now influences the rate $k_{21}$ (similar to the $k_{1a}$-modulation in the scheme 2-1a mentioned above), which yields scheme 3E-21. (iii) A coarse-grained model of DNA-looping in the \emph{lac}-operon \cite{Earnest2013}.
  {\bf Scheme 3M:}  A nucleosome and a specific, repressing TF compete for a promoter; changes in the input TF concentration correspond to 3M-a1, while changes in factors decreasing nucleosome occupancy correspond to 3M-1a.
  {\bf Scheme 3D:} (i) State $1$ is the closed chromatin formation, state $a$ is the empty promoter and in state $b$ a TF is bound to the DNA in such a way that it prohibits the closing of the chromatin (but still permits transciption) \cite{Mirny2010,SegalWidom2009rev}. In this way, even though the input molecule does not necessarily interact with the RNA polymerase directly, it can act as an activator (or rather as a de-repressor), yielding scheme 3D-ab. (ii) A scheme used to describe any promoter where the basal expression does not follow a Poisson process (optionally with different rates of expression from $a$ and $b$). (iii) Promoter with a TATA~box and a competing nucleosome (cf. 3E-21) if there is a significant amount of expression from the basal state $a$. (iv) A scheme used as a phenomenological model with unidentified third state to explain universal noise behavior in bacterial gene expression \cite{SanchezKondev2013rev, SoGolding2011}.
Cited references use similar models.}
  \label{fig:app_full}
  \end{center}
\end{figure*}

\section{Experimentally measured promoter switching rates}
Direct measurements of switching rates are rare since they require live imaging. Examples include the relative measurements of on-, off- and mRNA-production rates in \emph{E. coli} \cite{SoGolding2011} using the MS2-GFP system \cite{GoldingCox2005}, reporting $2-10$ fold higher on- than off-rates, and mRNA production rates an order of magnitude higher than the on-rates; original bursting reported in \cite{GoldingCox2005} finds the on-time duration to be roughly 6 and the off time 37 minutes in a synthetic \emph{E. coli} reporter system. Recently, on-rates of $\sim 3\cdot 10^{-2}\;\mathrm{min}^{-1}$, roughly ten-fold higher off-rates, and mRNA production rates ranging from $0-5\;\mathrm{min}^{-1}$ have been reported in mammalian cells using the luciferase reporter system \cite{SuterSchiblerNaef2011}. Using new high-throughput microfluidic methods, it is now possible to measure TF binding and unbinding times directly: Ref~\cite{GeertzMaerkl2012} reports mouse and yeast in vitro transcription factor dissociation rates between $\sim 10\;\mathrm{s}^{-1}$ and $10^{-2}\;\mathrm{s}^{-1}$, as well as the range of the corresponding association rates; it is, however, less clear if these can be unambiguously identified with switching rates in functional models.

A larger body of work extracts the rates of the two-state model from the noise characteristics (which are the primary measurement), \emph{assuming} the two-state model without diffusion noise is applicable. The reported Fano factors for mRNA counts vary, but are of the order of $1-10$. The typical values for kinetic parameters extracted for a range of \emph{E. coli} promoters are $10^{-3}-10^{-2}\;\mathrm{s}^{-1}$ for the on-rate, $10^{-1}-1\;\mathrm{s}^{-1}$ for the mRNA production rate when induced, and a variable off-rate that depends strongly on the induction level \cite{SoGolding2011}. Using a similar technique in mammalian cells, Raj et al \cite{Raj2006} extracted two-state parameters and found the on-rate normalized by mRNA decay time to be roughly of order unity, while the ratio of mRNA production rate to the off-rate varied from $\sim 10 -400$, depending on the system and the induction level. 

\section{Multi-state promoters as state-transition graphs}

In this section we describe the general method used to derive the behavior of noise and mean for different promoter architectures, followed by a calculation for one example architecture.

\subsection{Translating a state transition diagram into dynamic equations}
Let $\{a,b,\ldots M\}$ denote the states of the promoter that produce mRNA at a fixed rate $r$ and $\{1,2,\ldots N\}$ denote states without production. For $\mathcal S = \{a,b, \ldots M,1,2, \ldots N\} \ni i,j$, let $k_{ij}\geq 0, i\neq j$ be the rate with which the promoter switches from state $i$ to state $j$, $d$ be the rate of mRNA-degradation, and $p_i$ be the fractional occupancy of state $i$. For simplicity, we will only treat the case $M=1$ here.

{\bf Deterministic equations.}
The list of (non-zero) rates fully defines the state-transition graph, i.e. the promoter model. This directly translates into a linear system of equations that describes the dynamics of the system:
\begin{eqnarray}
   \partial_t \bm{p}  = \bm K \bm p \quad , \quad \text{with}
\end{eqnarray}
\begin{eqnarray}
   \bm K = \begin{bmatrix}
        -\sum_{j\in \mathcal S} k_{aj} & k_{1a} & \cdots & k_{Na}  \\
        k_{a1} & -\sum_{j\in \mathcal S} k_{1j} & \cdots & k_{N1}  \\
        \vdots & \vdots & \ddots & \vdots \\
        k_{aN} & k_{1N} & \cdots & -\sum_{j\in \mathcal S} k_{Nj}  \\
     \end{bmatrix}
      ,
\end{eqnarray}
$ \bm p = [ p_a, p_1 , \vdots , p_N]^T $,
subject to the normalization constraint $\sum_{i \in  \mathcal S} p_i = 1$.

The dynamics of mRNA are described by linking them to the activity of the promoter:
\begin{eqnarray}
   \partial _t m = r p_a - d m \quad .
\end{eqnarray}
To compute the average amount of mRNA $\bar m$ in steady state, we set the time derivatives to 0 and solve the linear set of equations
\begin{eqnarray}
 \bm K \bar{\bm p}= 0 \quad , \\
 \bar m = \frac{r}{d} p_a \quad .
\end{eqnarray}
As the occupancy of the active state $p_a$ is a function of the rates in $\bm K$, we can obtain the dependence of $\bar m$ on any rate of interest, i.e. we can obtain the regulation function.

{\bf Langevin approach to calculate noise behavior.}
For the noise behavior, we linearize Eqs~(4,5) of the main text around the mean:
\begin{eqnarray}
\bm p(t) = \bar{\bm p} + \delta p(t), \\
 m(t)=\bar m + \delta m(t)
\end{eqnarray}
and introduce the Fourier-transformed variables
\begin{eqnarray}
 \delta p_i(t) = (2\pi)^{-1}\int d\omega\,\delta \hat{p}_i(\omega)\exp{(-{\rm i}\omega t)}\quad ,\\
 \delta m(t) = (2\pi)^{-1}\int d\omega\,\delta \hat{m}(\omega)\exp{(-{\rm i}\omega t)} \quad ,
\end{eqnarray}
so that we get the linear response to random fluctuations:
\begin{eqnarray}
  (-{\rm i} \omega) \delta  \hat{\bm p} = \bm K \delta \hat{\bm p} +  \hat{\bm \xi}\quad ,\label{eqn:lgv15}\\
  (-{\rm i}\omega) \delta \hat{m} = r \delta \hat{p}_a - d \delta \hat{m} + \hat{\xi}_m\quad . \label{eqn:lgv1}
\end{eqnarray}
The statistics of the Langevin forces are given by:
\begin{eqnarray}
  \langle \hat{\xi}_i^* \hat{\xi}_j \rangle = -(\hat p_i K_{ij} + \hat p_j K_{ji})\quad ,\label{eqn:lgv17}\\
  \langle \hat{\xi}_m^* \hat{\xi}_m \rangle = 2 d \bar m \quad ; \label{eqn:lgv2}
\end{eqnarray}
to see this for the variances, consider $\langle \xi_i \xi_i^* \rangle = -2\hat p_i K_{ii}  = 2 \hat p_i \sum_jk_{ij}$, since all entries in the diagonal of $\bm K$ are negative. This is two times the rate of leaving state $i$. Similarly, for $\langle \xi_m^* \xi_m \rangle$ the variance is two times the rate of degrading a molecule. The factor of two comes from the fact that we consider a system at steady state, so the rates of entering and leaving a state (or creating an destroying a molecule) must be equal.
For the covariances $ \langle \xi_i^* \xi_j \rangle $ ($i\neq j$), the two Langevin forces are anti-correlated, since leaving one state means entering another. The rate of changing between the two states is the probability of being in state $i$ ($p_i$) times the rate of transition from that state into the other ($k_{ij}=K_{ij}$) -- and this holds for both directions between the pair of states. Also, since we assume that production of mRNA and promoter switching are independent, $ \langle \xi_i^* \xi_m \rangle= 0$ for all states $i$.

To get the variance in mRNA, we compute
$\sigma^2_m = (2\pi)^{-1} \int d\omega\, |\delta\hat{m}(\omega)|^2$, where $\delta\hat{m}(\omega)$ is obtained from Eq~(\ref{eqn:lgv1}) as
\begin{eqnarray}
\langle \delta \hat{m}^* \delta \hat m\rangle = \frac{2 d \bar m}{d^2+\omega^2} + \frac{r ^2}{d^2+\omega^2}\langle \delta \hat p_a ^* \delta \hat p_a \rangle \quad , \label{eqn:lgvmm}
\end{eqnarray}
where $ \langle \delta \hat p_a ^* \delta \hat p_a \rangle $ is calculated by solving Eq~(\ref{eqn:lgv15}) and using the Langevin noise magnitudes from Eqs~(\ref{eqn:lgv17},\ref{eqn:lgv2}).

With the assumption $d\ll k_{ij}$, Eq~(\ref{eqn:lgv1}) becomes
\begin{eqnarray}
0 = \bm K (\delta \bm \hat{p}) + \bm \hat{\xi}  \quad ,\\
\sum_i \delta \hat p_i = 0 \quad ,
\end{eqnarray}
which simplifies the expressions for the $\delta \hat p_i$.
This is because the terms with $(-i\omega)$ in the denominator (as seen in the next section in Eqs~(\ref{eq:omega1},\ref{eq:omega2})) would give an additional, multiplicative term of the form $1/(k_{ij}^2+\omega^2)$ in Eq~(\ref{eqn:lgvmm}). The $\omega$-dependence of these terms can be neglected for the integration, since for $d\ll k_{ij}$ we have
\begin{eqnarray}
\frac{1}{k_{ij}^2+\omega^2}\frac{1}{d^2+\omega^2} \approx \frac{1}{k_{ij}^2}\frac{1}{d^2+\omega^2} \quad .
\end{eqnarray}

\subsection{Example: Dual regulation (3E-1a)}
We are interested in a system where the promoter of a gene can either be occupied by an activator (present at concentration $a$) or a repressor (present at concentration $b$).
If it is in the active state, it produces mRNA at a constant rate $r$, which is later degraded at rate $d$.

\begin{center}
\begin{tikzpicture}[->,>=stealth',shorten >=1pt,auto,node distance=1.8cm,
                    semithick]
  \tikzstyle{every state}=[fill=white,draw=black,text=red]

  \node[state]         (E)                                                {$1$};
  \node[state,double]         (A) [right of=E, fill=none, text=black]     {$a$};
  \node[state]         (B) [left of=E]                             {$2$};
  \node[state]         (M) [right of=A, fill=none, draw=none, text=blue]  {$m$};
  \node[state]         (O) [right of=M, fill=none, draw=none, text=black] {$O$};

  \path (E) edge [bend left]  node               {$a\, k_{+}$} (A)
            edge [bend right] node [above]  {$b\, k_{+}$} (B)
        (A) edge              node               {$k_{-}^a$}  (E)
            edge [dashed]  node               {$r$}        (M)
        (B) edge              node [below] {$k_{-}^b$}  (E)
        (M) edge [dotted]        node               {$d$}    (O);
\end{tikzpicture}
\end{center}

{\bf Deterministic equations.}
Following the setup from the last section, we translate the state transition diagram into a matrix that describes the dynamics at the promoter:
\begin{equation}
   \begin{bmatrix}
       \partial _t p_a  \\
       \partial _t p_1  \\
       \partial _t p_2  \\
   \end{bmatrix}
   =
   \begin{bmatrix}
        -k_{-}^a & a k_{+} & 0  \\
        k_{-}^a & - (a k_{+}+b k_{+}) & k_{-}^b  \\
        0 & b k_{+} & -k_{-}^b  \\
   \end{bmatrix}
      \cdot
   \begin{bmatrix}
        p_a  \\
        p_1  \\
        p_2  \\
   \end{bmatrix} \quad .
\end{equation}
This is then the basis for a description of the dynamics of the output (here mRNA):
\begin{eqnarray}
\partial _t\, m = r p_a - d m \quad ,\\
\partial _t\, p_a = a k_{+} p_1 - k_-^a p_a \quad ,\\
\partial _t\, p_2 = b k_+ p_1 - k_-^b p_2 \quad ,\\
p_a + p_1 + p_2 = 1 \quad .
\end{eqnarray}

With the definitions $A=\frac{a k_+}{k_-^a}$, $B=\frac{b k_+}{k_-^b}$, $S=1+A+B$ and $R=\frac{r}{d}$ we get for the steady state:
\begin{eqnarray}
\bar p_a = A/S, \ \bar p_1 = 1/S, \ \bar p_2 = B/S \quad ,\\
\bar m = R \bar p_a = RA/S \quad .
\end{eqnarray}

{\bf Langevin approach.}
To see how the dynamics of the promoter influence the statistics of mRNA we perturb the systems with Langevin forces (while still keeping the gene copy number constant):
\begin{eqnarray}
\partial _t\, m = r p_a - d m + \xi_m  \quad ,\\
\partial _t\, p_a = a k_{+} p_1 - k_-^a p_a + \xi_a \quad ,\\
\partial _t\, p_2 = b k_+ p_1 - k_-^b p_2 + \xi_2 \quad ,\\
p_a + p_1 + p_2 = 1 \quad .
\end{eqnarray}
The mean of the Langevin forces is zero ($\langle \xi_i (t) \rangle = 0$) and they are uncorrelated in time ($\langle \xi_i (t) \xi_i (t')\rangle \propto \delta(t-t')$).

We linearize around the mean, where deviations from the mean are denoted by $\delta$:
\begin{eqnarray}
m(t) = \bar m + \delta m(t) \quad ,\\
p_a(t) = \bar p_a + \delta p_a(t)  \quad ,\\
p_2(t) = \bar p_2 + \delta p_2(t) \quad ,\\
\delta p_1 = - \delta p_a - \delta p_2 \quad .
\end{eqnarray}
After inserting the linearized equations into the Langevin approach we perform a Fourier transform:
\begin{eqnarray}
-i\omega \delta \hat m = r \delta \hat p_a - d \delta \hat m + \hat \xi_m \label{eq:m}  \quad ,\\
-i\omega \delta \hat p_a(\omega) = a k_+ (-\delta \hat p_2 - \delta \hat p_a) - k_-^a \delta \hat p_a + \hat \xi_a \label{eq:lin1}  \quad ,\\ 
-i\omega \delta \hat p_2(\omega) = b k_+ (-\delta \hat p_a - \delta \hat p_2) - k_-^b \delta \hat p_2 + \hat \xi_2 \label{eq:lin2}  \quad .
\end{eqnarray}
Starting with the equations for the occupancies, we rewrite Eqs~(\ref{eq:lin1},\ref{eq:lin2}) and use the approximation that $d$ is significantly slower than the other rates to get:
\begin{eqnarray}
\delta \hat p_a(\omega) = \frac{a k_+ \delta \hat p_1 +\hat \xi_a}{k_-^a-i\omega}\approx
A \delta \hat p_1+\frac{\hat \xi_a}{k_-^a} \label{eq:omega1} &  \quad , \\ 
\delta \hat p_2(\omega) = \frac{b k_+ \delta \hat p_1 +\hat \xi_2}{k_-^b-i\omega}\approx
B \delta \hat p_1+\frac{\hat \xi_2}{k_-^b} \label{eq:omega2} & \quad , \quad \text{or}\\
\delta \hat p_a = -\delta \hat p_2 \frac{A}{(1+A)} +\frac{\hat \xi_a}{k_-^a}\frac{1}{(1+A)}  \quad ,\\ 
\delta \hat p_2 = -\delta \hat p_a \frac{B}{(1+B)} +\frac{\hat \xi_2 }{k_-^b}\frac{1}{(1+B)}  \quad .
\end{eqnarray}
Solving this system yields:
\begin{eqnarray}
\delta \hat p_a =  - \frac{\hat \xi_2 }{k_-^b}\bar p_a +\frac{\hat \xi_a}{k_-^a}(\bar p_1+ \bar p_2) \label{eq:pa}  \quad .
\end{eqnarray}
The variances of the Langevin forces are:
\begin{eqnarray}
\langle \hat \xi_a^* \hat \xi_a\rangle = 2 k_-^a\bar p_a \quad , \\
\langle \hat \xi_2^* \hat \xi_2\rangle = 2 k_-^b\bar p_2 \quad , \\
\langle \hat \xi_m^* \hat \xi_m\rangle = 2 d\bar m \quad ,
\end{eqnarray}
and their covariances vanish, since the direct transition from state $a$ to state $2$ is not allowed.
From Eqs (\ref{eq:m},\ref{eq:pa}) we get:
\begin{eqnarray}
\langle \delta \hat p_a ^* \delta \hat p_a \rangle = 2\frac{\bar p_2}{k_-^b}\bar p_a^2 + 2  \frac{\bar p_a}{k_-^a}(\bar p_1+\bar p_2)^2  \quad ,\\
\langle \delta \hat m^* \delta \hat m\rangle = \frac{2 d \bar m}{d^2+\omega^2} + \frac{r ^2}{d^2+\omega^2}\langle \delta \hat p_a ^* \delta \hat p_a \rangle \quad . \label{eq:omega3}
\end{eqnarray}
Finally, with $ \frac{1}{2\pi} \int_{-\infty}^{\infty}2\frac{1}{x^2+\omega^2} \, d\omega = \frac{1}{x} $ we get:
\begin{eqnarray}
\sigma_m^2 =
 \frac{ d \bar m}{d} + \frac{r ^2}{d}\left(\frac{p_2}{k_-^b}p_a^2 + \frac{p_a}{k_-^a}(p_1+p_2)^2 \right) = \nonumber \\=
\bar m \left[1 + r \left(\frac{p_2}{k_-^b}p_a + \frac{1}{k_-^a}(1-p_a)^2 \right)\right] \label{eq:3e1afinal} \quad .
\end{eqnarray}
This is one description of noise in the 3E architecture. To get the noise characteristics for modulation scheme 3E-1a, we need to express $p_2$ in terms of $p_a$ (not shown).
From Eq~(\ref{eq:3e1afinal}) we can see that in the absence of repressors ($p_2=0$) and also for very fast unbinding of the repressors ($k_-^b\to \infty$) the noise shows the quadratic dependence on the occupation of the promoter that we see in the corresponding two-state model 2-1a.

\begin{figure}[!h]
  \begin{center}
  \includegraphics*[width=1.65in]{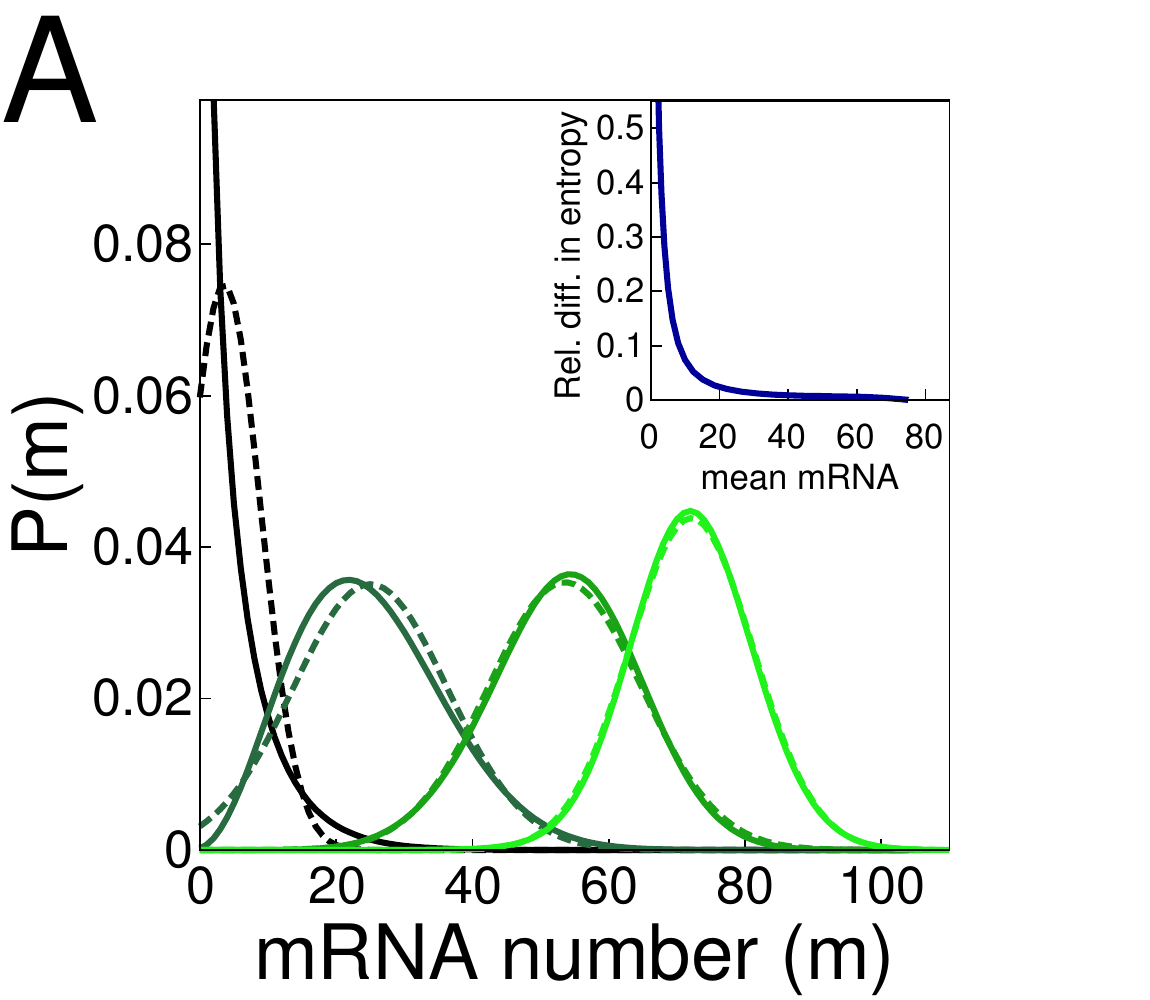}
    \includegraphics*[width=1.65in]{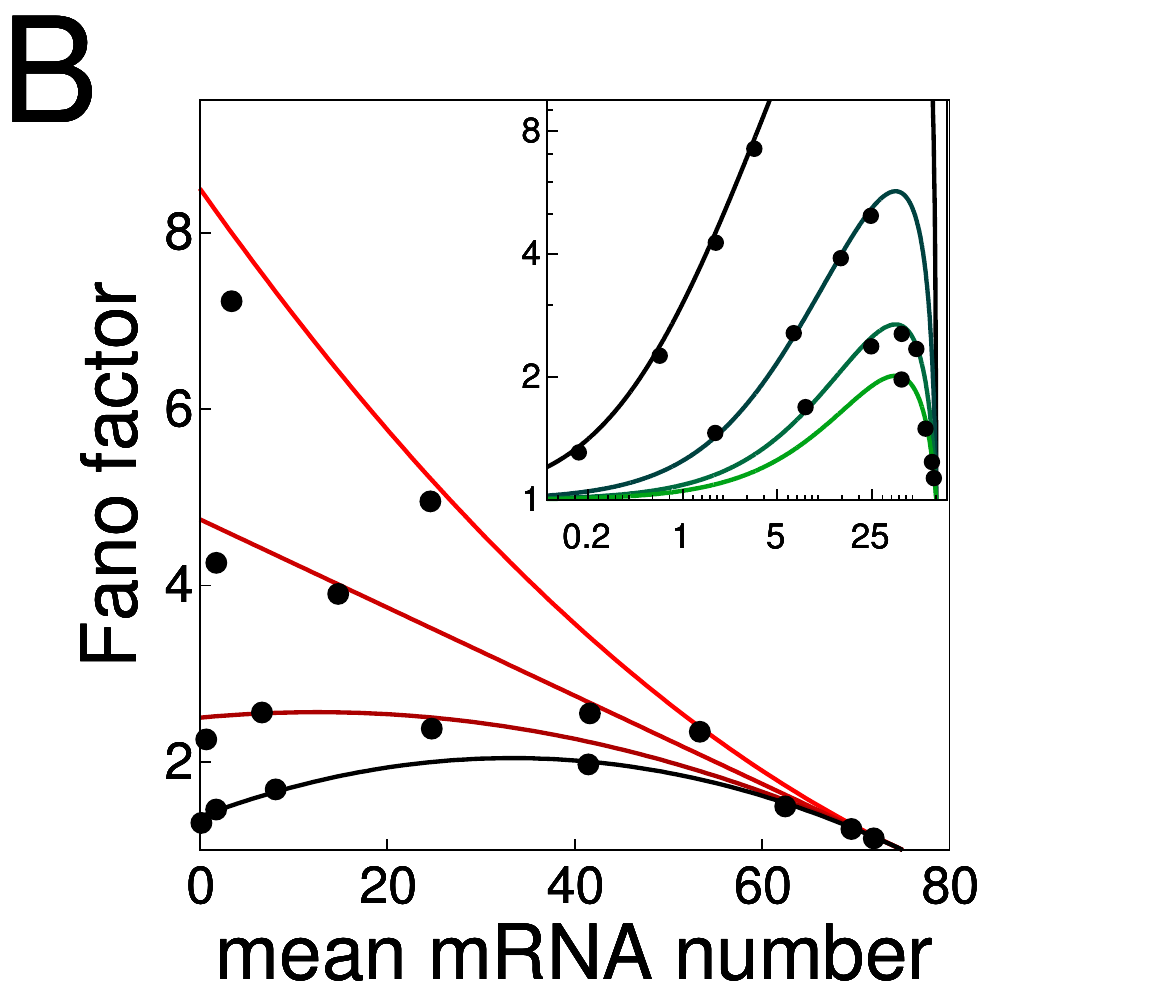}
  \caption{{\bf Comparison of Langevin approach to other methods.} {\bf (A)} Probability distribution of the model 3E for different values of $a k_+$ (thus treating it as 3E-1a). Solid lines are the numerical solution of the master equation and dashed lines are a gaussian approximation using the analytical expressions for mean and variance described in the main text. Different colors are for different values of $ak_+$. Parameters for this plot: $a k_+=[1,10,50,500]$, $k_-^a=10$, $b k_+=10$, $k_-^b=10$, $r=75$ and $d=1$.
  Inset: The relative difference in entropy between the full distribution and the gaussian approximation used in this study (i.e. using the results for mean and variance from the Langevin method). The error drops to $<5 \% $ very fast, with the main difference at slow rates stemming mainly from non-gaussianity at low expression levels. Importantly, for information calculations this non-gaussianity is mitigated by the protein averaging; even without the averaging, the effect on the comparison between architectures is minor. {\bf (B)} Comparison of Fano factors for different mean expression levels for the 3E-1a model (black data points  = Gillespie simulation, solid lines = Langevin method). Different colors correspond to different values for $k_-^a$. For small values of $k_-^a$ and $ak_+$ we start to see deviations since the approximation $d\ll k_{ij}$ no longer holds. Parameters for the Gillespie simulations: $k_-^a = [10,20,50,200]$, $ak_+ = [1,10,50,500]$, $b k_+=10$, $k_-^b=10$, $r=75$ and $d=1$. The Fano factor was calculated from 10000 runs. Error bars from resampling are smaller than the symbols.
  Inset: The same results from the Gillespie simulations replotted (on a log-log scale) and compared to the noise characteristics of the 3E-a1 scheme, i.e. the rate $k_-^a$ is modulated to obtain different mean mRNA levels. The solid lines are the noise characteristics calculated with the Langevin methods for different (fixed) values of $ak_+ = [1,10,50,500]$.
  }
  \label{fig:num_vs_ana}
  \end{center}
\end{figure}

\subsection{Comparison to other methods}
The results obtained with the Langevin approach were compared against two other methods: \emph{(i)} the exact numerical solution of the chemical master equation and \emph{(ii)} results from stochastic simulation using the Gillespie algorithm.
 Two kinds of comparisons are relevant: first, how well the gaussian distribution approximates the true distribution of mRNA levels; and second,  how the Langevin-derived expressions for the noise characteristics compare to the exact values.

Fig~\ref{fig:num_vs_ana}A compares the distribution of mRNA levels obtained from the numerical solution of the chemical master equation to the gaussian approximation for the dual regulation architecture discussed in the last section.

The stochastic simulation algorithm is time consuming and offers no special benefit for the simple systems studied here, but we have nevertheless checked a few example architectures against simulation results. The results for dual regulation are shown in Fig~\ref{fig:num_vs_ana}B.
Values for $ak_+$ and $k_-^a$ were chosen from a grid. This makes it possible to show the agreement with the Langevin-derived noise characteristics in two different modulation schemes (cf. inset in Fig~\ref{fig:num_vs_ana}B).

Another way to obtain analytical expressions for the mean and variance of the mRNA-distributions is the method of partial moments (e.g., \cite{SanchezKondev2011, SanchezKondev2008}). 
While this method can also be used to derive higher moments, a minor advantage of the Langevin method for the purposes here is that the approximation $d\ll k_{ij}$ can be used earlier in the derivations, leading to simpler expressions.




\end{document}